\documentclass[pra,
a4paper,
showpacs,
twocolumn,
superscriptaddress]{revtex4-1}
\usepackage[T2A]{fontenc}
\usepackage[utf8]{inputenc}
\usepackage[english]{babel}
\usepackage{amssymb}
\usepackage{amsmath}
\usepackage{amsfonts}
\usepackage{braket}
\usepackage{graphicx}
\usepackage{bm}
\usepackage{color}
\usepackage{multirow}
\usepackage{natbib}
\usepackage{hyperref}
\usepackage[T2A]{fontenc}
\usepackage{verbatim}
\usepackage{dcolumn}
\usepackage{ulem}
\usepackage{float}
\usepackage[utf8]{inputenc}   
\usepackage[T2A]{fontenc}     

\usepackage[dvipsnames]{xcolor}
\usepackage{graphicx}% Include figure files
\def \be {\begin{equation}}
\def \ee {\end{equation}}
\def \bea {\begin{align}}
\def \eea {\end{align}}
\def \p {\partial}
\def \BEA {\begin{eqnarray}}
\def \EEA {\end{eqnarray}}

\def \BC {\begin{cases}}
\def \EC {\end{cases}}

\begin{document}
\title
{ 
Qubit decoherence in dissipative two-photon resonator: real-time instantons and Wigner function
}

\author{V.\,Yu. Mylnikov}
\address{Ioffe Institute,
194021 St.~Petersburg, Russia}
\author{S.\,O.~Potashin}
\address{Ioffe Institute,
194021 St.~Petersburg, Russia}
\author{A. Kamenev}
\address{School of Physics and Astronomy, University of Minnesota, Minneapolis, MN 55455, USA  }
\address{William I. Fine Theoretical Physics Institute, University of Minnesota, Minneapolis, MN 55455, USA }

%\date{\today}
%\pacs{72.80.Vp, 73.23.Ad, 73.63.Bd}

\begin{abstract} 
We study the quantum dynamics of a single bosonic cavity subject to two-photon driving and two-photon dissipation in the presence of finite detuning. Exploiting a hidden time-reversal symmetry, the Wigner representation and the WKB method, we introduce an effective phase-space potential for description of the steady state. It reveals two attracting points, which are metastable due to quantum fluctuations. By employing the Keldysh real-time path integral formalism, we compute the instanton trajectory governing the quantum activation process between these attractors and establish a fundamental connection with the Wigner representation. This relation unifies the steady-state phase-space description with dynamical quantum activation processes. We also derive an analytical expression for the decoherence rate of the system. Our work provides a coherent theoretical framework for analyzing quantum bistability, metastability, and decoherence in driven-dissipative nonlinear resonators, with direct implications for the design of bosonic qubits and quantum information processing.\end{abstract}

\maketitle

\section{Introduction}
The open quantum systems subject to coherent driving and dissipation exhibit a rich landscape of non-equilibrium phenomena, including dissipative phase transitions \cite{Carmichael2015,Bartolo2016,Minganti2016,Mylnikov2022,Mylnikov2025}, metastability \cite{Meaney2014}, and the formation of non-classical states of light \cite{Gilles1994,Brunelli2018}. In particular, cavities with nonlinear interactions, such as those engineered in superconducting circuits \cite{Leghtas2015}, provide a versatile platform for exploring quantum dynamics in regimes where driving, detuning, and dissipation compete. Among such systems, the two-photon driven-dissipative resonator \cite{Mirrahimi2014} stands out as a paradigmatic model for studying quantum bistability \cite{Bartolo2016,Roberts2020}. This approach finds a key application in quantum computing via bosonic codes \cite{Berdou2023,Reglade2024}. Theoretical descriptions of these systems must account for both coherent evolution and dissipative processes, typically captured by the Lindblad master equations \cite{Kamenev,Walls,Sieberer2016}. Although semiclassical approximations can reveal classical fixed points and bifurcations, a full quantum treatment is essential to understand phenomena such as quantum tunneling and the lifetime of metastable states.

Two frameworks have been widely employed: the phase-space approach via the Wigner distribution, which offers an intuitive representation of quantum states in terms of quasi-probability densities, and the real-time path integral formalism, which enables the calculation of dynamical tunneling rates via instanton methods. However, a coherent and explicit connection between these two approaches, especially in the presence of detuning and nonlinear dissipation, has remained underexplored.

In this work, we address this gap by studying a single bosonic cavity driven by a two-photon pump and subject to two-photon dissipation. The interplay between the pump and dissipation leads to a bistable phase space with two symmetry-related attractors. Quantum fluctuations induce the activation process between these attractors. Our primary objectives are threefold: to derive the exact stationary Wigner function for the detuned two-photon cavity by exploiting a recently identified \textit{hidden time-reversal symmetry} \cite{Roberts2021}, which constrains its functional form and simplifies the underlying equation of motion. Construct an effective phase-space potential within the WKB approximation, which provides an intuitive picture of the quantum metastability and serves as a bridge to dynamical calculations. Establish a direct and explicit connection between the Wigner representation and the real-time instanton formalism and show that the quantum field along the instanton trajectory is proportional to the gradient of effective potential.

We develop a phase-space effective potential within the WKB framework to provide an intuitive, semi-classical picture of quantum metastability. This potential acts as a crucial bridge to real-time dynamics. We then establish a direct correspondence between this Wigner-representation formalism and the real-time instanton method. A key result of this connection is the demonstration that the quantum field evolution along the instanton trajectory is governed by the gradient of this effective potential. By unifying these perspectives, we not only recover known results for the resonant case, but also extend the analysis to finite detuning, where the phase space becomes genuinely complex and the instanton dynamics more intricate. We derive a closed-form expression for the quantum decoherence rate, highlighting how detuning modifies the tunneling exponent. Our approach demonstrates that the stationary Wigner function encodes sufficient information to predict dynamical tunneling rates. This result underscores the deep interplay between steady-state quantum distributions and real-time fluctuation phenomena. 

This paper is structured as follows. In Section II, we introduce the model and its Lindblad master equation. Section III is devoted to the Wigner representation, where we derive the exact stationary solution and its WKB approximation. In Section IV, we formulate the real-time instanton approach within the Keldysh framework, compute the instanton trajectory, and link it to the Wigner effective potential. Finally, in Section V, we present our conclusions and discuss broader implications for driven-dissipative quantum systems. Our work provides a unified theoretical framework for analyzing quantum metastability in nonlinear photonic systems, with potential applications in quantum information processing.
 
\section{The Model}
We consider a single bosonic cavity driven coherently by a two-photon pump and subject to two-photon dissipation. Its evolution is described by the Lindblad master equation (with $\hbar = 1$):
\begin{equation}\label{Lindblad}
\partial_{t}\hat{\rho}
= -i[\hat{H},\hat{\rho}]
+ \eta\,\mathcal{D}_{\hat{a}^{2}}(\hat{\rho}),
\end{equation}
where $\mathcal{D}_{\hat{X}}(\hat{\rho}) = \hat{X}\hat{\rho}\hat{X}^{\dagger}
- \{\hat{X}^{\dagger}\hat{X},\hat{\rho}\}/2$ is the dissipative superoperator. In the rotating frame of the pump field with frequency $\omega_{p}$, the Hamiltonian takes the time-independent form:
\begin{equation}
\hat{H}
= -\Delta\,\hat{a}^{\dagger}\hat{a}
+ \frac{iG}{2}\left(\hat{a}^{\dagger 2} - \hat{a}^{2}\right),
\end{equation}
where the bosonic operators $\hat{a}$ and $\hat{a}^{\dagger}$ satisfy the canonical commutation relation $[\hat{a},\hat{a}^{\dagger}] = 1$. Here, $\Delta = \omega_{p} - \omega_{c}$ is the detuning between the pump and cavity frequencies and $G$ is the amplitude of the coherent two-photon drive. The two-photon dissipation with rate $\eta$ is described by the quantum jump operator $\hat{a}^{2}$. Such processes can be implemented, for example, using Josephson-junction-based parametric coupling between superconducting resonators~\cite{Leghtas2015}. This model therefore captures the interplay between cavity detuning, two-photon driving, and the corresponding two-photon loss. This simple system demonstrates the key features of quantum bistability and offers unexpected and nontrivial physics \cite{Mirrahimi2014,Leghtas2015,Bartolo2016,Roberts2020,Mylnikov2022}. In the following Section we investigate steady state and time dynamics of the system using the Wigner representation and real-time instanton formalism. 
\section{The Wigner function }

In this Section, we study the steady-state of the system using the Wigner distribution function \cite{Cahill1969,Scully}. Starting from Eq. \eqref{Lindblad} an equivalent equation of motion for the Wigner function can be derived \cite{Scully}:
\begin{widetext}
\begin{equation}\label{WignerEq}
\begin{aligned}
& \frac{\partial W}{\partial t} = -\left[\frac{\partial}{\partial\alpha} (i\Delta \alpha + G\alpha^* - \eta\alpha(|\alpha|^2-1)) + c.c.\right] W+ 2\eta \frac{\partial^2}{\partial\alpha \partial\alpha^*} \left(|\alpha|^2 - \frac{1}{2}\right) W + \frac{\eta}{4} \left(\frac{\partial^3}{\partial\alpha \partial(\alpha^*)^2} \alpha^* + c.c.\right) W.
\end{aligned}
\end{equation}
\end{widetext}
The first term in Eq. \eqref{WignerEq} represents  the classical drift in phase space under the action of external forces, while the second term accounts for diffusion. The third term, which contains a third-order derivative, does not have a classical analog in the Fokker-Planck equation. This term is a fundamental signature of non-classicality and significantly complicates the analysis of both the steady-state properties and dynamics of the system. The common alternative is to adopt a complex P-distribution that satisfies the corresponding Fokker-Planck-like equation of motion. This approach facilitates the determination of the stationary distribution function \cite{Bartolo2016} and can be applied to calculate the decoherence rate \cite{Kinsler1991,Mylnikov2025II}. However, the P-representation has several undesirable features that complicate its use. First, unlike the Wigner function, which depends on a single complex variable $\alpha$, it is a function of two complex variables, thus doubling the dimensionality of the effective phase space. Second, the P-representation is not unique \cite{Drummond1980}. Third, stationary P-distribution can exhibit singularities, challenging its interpretation as a well-behaved probability density \cite{Drummond1980,Bartolo2016}. However, these singularities can be handled by choosing appropriate integration paths in the complex phase space \cite{Bartolo2016}. The Wigner function inherently avoids all three of these problems. A common approach to simplifying Eq. \eqref{WignerEq} is the truncated Wigner approximation \cite{Walls, Kinsler1991}, which involves neglecting the third-order derivatives in Eq. \eqref{WignerEq}. However, it significantly overestimates the decoherence rate \cite{Kinsler1991} and thus cannot provide a correct description of the complete quantum problem. However, the exact stationary Wigner function can be obtained in a very indirect way using the P-representation \cite{Bartolo2016} and the quantum absorber method \cite{Roberts2020}.  
\begin{figure*}
\includegraphics[width=0.65\textwidth]{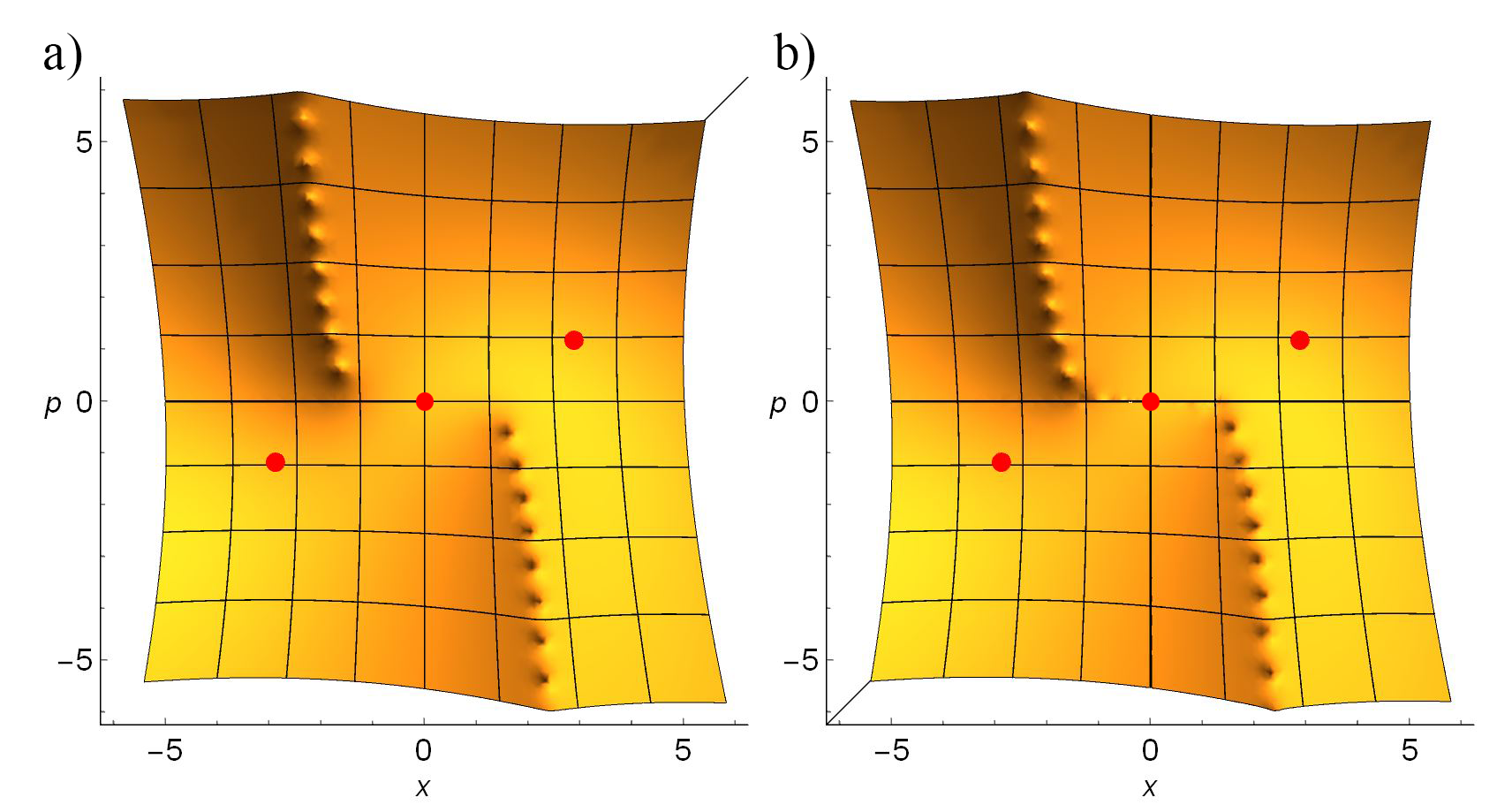}
\caption{(a,b) The minus logarithm of the stationary Wigner function, $-\ln(W_0)$, vs the photonic quadratures $x=\sqrt{2}\rm{Re}(\alpha)$ and $p=\sqrt{2}\rm{Im}(\alpha)$. It is obtained from (a) the exact solution \eqref{PsiExact} and (b) the WKB method \eqref{WKBFinal}.  The red dots highlights location of the semi-classical fixed point \eqref{semicl} . The system's parameters are set to $G=10, \Delta=7,\eta=1$.}
\label{FigLogW}
\end{figure*}

In this Section, we show how to obtain the stationary Wigner function directly from Eq. \eqref{WignerEq}. The main simplification comes from the fact that the detuned two-photon quantum cavity obeys the recently introduced hidden-time reversal symmetry \cite{Roberts2021}. As shown below, the stationary Wigner function for such systems has a very simple form:
\begin{equation}\label{Decomp}
W_0(\alpha,\alpha^*) = e^{-2\alpha^*\alpha} \Psi_1(\alpha) \Psi_2(\alpha^*).
\end{equation}
where the functions $\Psi_{1}(\alpha)$ and $\Psi_{2}(\alpha^*)$ depend on only one complex variable. This form of the stationary solution can be guessed in several ways. First, the stationary Wigner function can be calculated from the stationary P-distribution function \cite{Bartolo2016} and is shown to obey Eq. \eqref{Decomp}. Second, one can apply the coherent quantum-absorber method \cite{Roberts2020}, which results in the Wigner function with the same property \eqref{Decomp}. 

Here, we demonstrate that one can start from the decomposition \eqref{Decomp} and Eq. \eqref{WignerEq} to determine the steady state properties of the system. Let us substitute Eq. \eqref{Decomp} into Eq. \eqref{WignerEq} and obtain the following equations on $\Psi_{1}(\alpha)$ and $\Psi_{2}(\alpha^*)$ (Appendix \ref{AppPsi}): 
\begin{equation}\label{difeq1}
\begin{gathered}
\alpha \eta \Psi_1''(\alpha) + 2 i \Delta \Psi_1'(\alpha) - 4 \alpha G \Psi_1(\alpha) = 0,
\end{gathered}
\end{equation}
\begin{equation}\label{difeq2}
\begin{gathered}
\alpha^*   \eta  \Psi_2''(\alpha^* )-2 i \Delta  \Psi_2'(\alpha^*  )-4 \alpha^*  G \Psi_2(\alpha^*  )=0.
\end{gathered}
\end{equation}
The \eqref{difeq1} and \eqref{difeq2} are second-order differential equations of one variable, which have a much simpler form compared to the starting Eq. \eqref{WignerEq}. A very similar equation appears in the paper \cite{Roberts2020}. However, it has different physical meanings of the Segal-Bargmann representations of the pure-state wave function in the coherent quantum-absorber method. Here, we derived it in the context of the Wigner distribution function. We will also assume that $\Psi_2(\alpha^* )=\Psi_1(\alpha)^*$, because the equation \eqref{difeq2} is a complex conjugation of the equation \eqref{difeq1}. The exact regular solution of Eq. \eqref{difeq1} is as follows: 
\begin{equation}\label{PsiExact}
\Psi_1(\alpha) =\mathcal{N} \exp\left({2 
   \sqrt{\frac{G}{\eta }}\alpha}\right) \, _1F_1\left(\frac{i \Delta
   }{\eta };\frac{2 i \Delta
   }{\eta };-4 \alpha 
   \sqrt{\frac{G}{\eta
   }}\right),
\end{equation}
where $\mathcal{N} $ is a normalization coefficient and $_1F_1$ denotes a Kummer’s hypergeometric function. After the substitution of Eq. \eqref{PsiExact} into the Wigner function \eqref{WignerEq} one can obtain the exact stationary steady state, previously derived in the framework of complex P-representation \cite{Bartolo2016} or the quantum-absorber method \cite{Roberts2020}. 

Nevertheless, we are interested in the regime of small nonlinear dissipation $\eta\ll G$. As we show below, one can apply standard WKB methods to Eq. \eqref{difeq1} to obtain a much simpler expression for the stationary Wigner function. In addition, it will allow us to demonstrate the deep connection between the Wigner function approach and the instanton calculations, which were discussed in a following Section. As a result, we obtain the following expression (Appendix \ref{AppWKB}): 
\begin{equation}
\Psi^{\pm}_1(\alpha) = A_{\pm}(\alpha)\exp(\phi_{0,\pm}(\alpha)/\eta),
\end{equation}
where the exponential function $\phi_{0,\pm}$ is given by:
\begin{equation}\label{phi0pm}
\begin{gathered}
\phi_{0,\pm}(\alpha)=\\=\mp i\sqrt{\Delta ^2-4 \eta G\alpha ^2}-i\Delta  \ln \left( \frac{\sqrt{\Delta ^2-4 \eta G\alpha ^2}\mp\Delta}{\eta}\right),
\end{gathered}
\end{equation}
and $A_{\pm}(\alpha)$ is a slowly changing pre-exponential factor. To match the WKB solutions with the exact function \eqref{PsiExact} we find their asymptotics at infinity and obtain the following final expression (Appendix \ref{App2Sol}):  
\begin{equation}\label{WKBFinal}
\begin{gathered}
\Psi_1(\alpha) =\\
=\mathcal{N} 2^{-i\Delta / \eta} \frac{\Gamma (2i\Delta / \eta)}{\Gamma (i\Delta / \eta)}
(e^{-\pi\frac{\Delta}{\eta} }\Psi^{+}_1(\alpha)+e^{\pi\frac{\Delta}{\eta} }\Psi^{-}_1(\alpha)).
\end{gathered}
\end{equation}
A good comparison between the exact and WKB solutions is shown in Fig. \ref{FigLogW}, where the "effective potential" $-\ln(W)$ is depicted as a function of the photonic quadratures. One can see that it has two global minimums at and one saddle point near zero. The minus solution is dominant in the first and third quadrants of the $(x, p)$, plane, while the plus solution dominates in the second and fourth quadrants. The switching between the two solutions occurs when the real parts of the exponents coincide, i.e., when$\mathrm{Re}(\phi_{0,+}-\phi_{0,-})=0$. This condition gives rise to the "mountain range" structure observed in Fig. \ref{FigLogW}, located in the second and fourth quadrants. The peaks along this switching line can be understood as interference effects between the two solutions in Eq. \eqref{WKBFinal}. Although the plus and minus solutions share the same amplitude, they accumulate different phases. Their interference can cause the Wigner function to vanish, leading to a divergence in its logarithm. 

\begin{figure}
\includegraphics[width=0.35\textwidth]{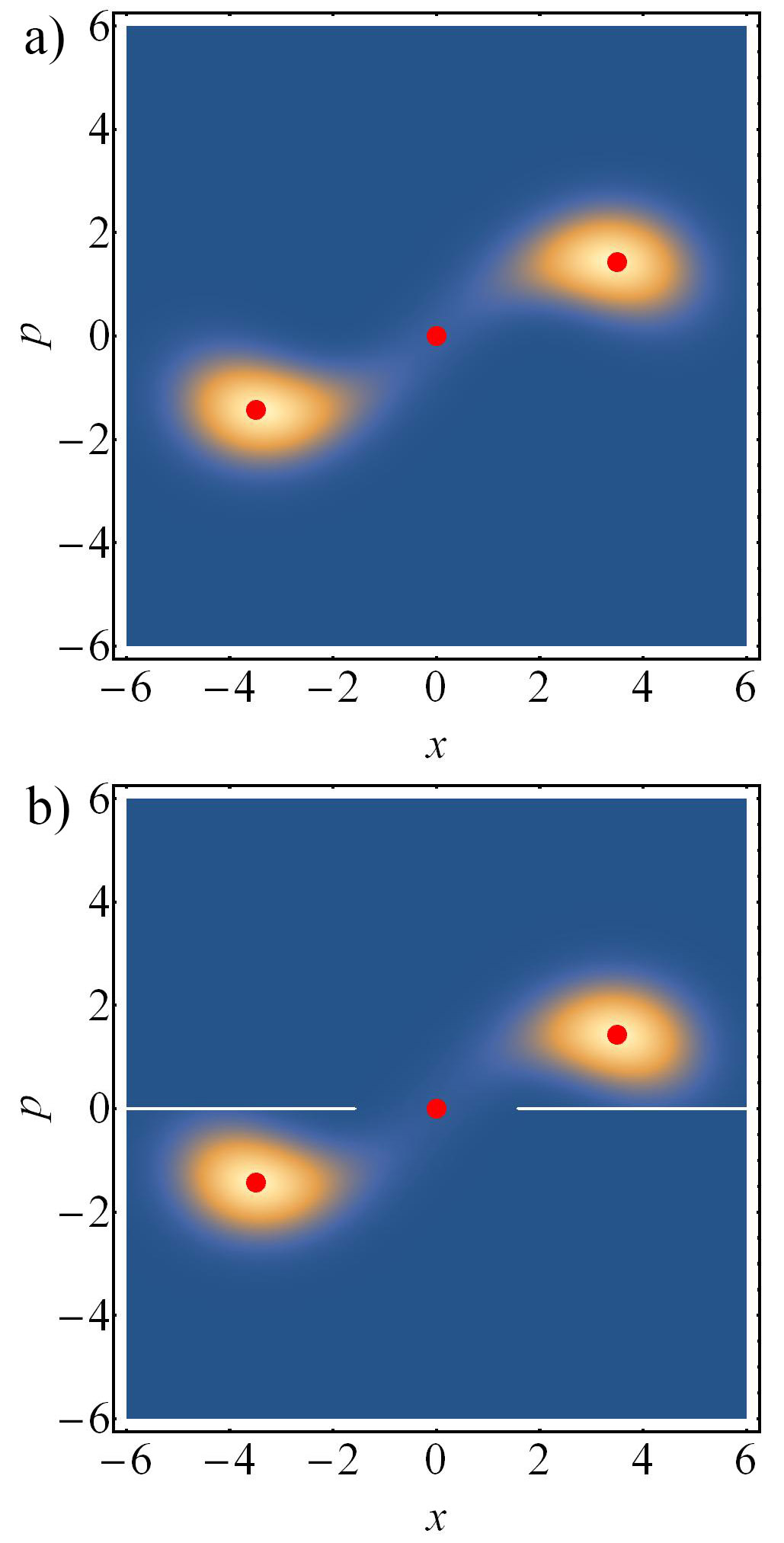}
\caption{(a,b) The Wigner distribution vs the photonic quadratures $x=\sqrt{2}\rm{Re}(\alpha)$ and $p=\sqrt{2}\rm{Im}(\alpha)$. It is obtained from (a) the exact solution \eqref{PsiExact}-\eqref{Decomp} and (b) the effective potential approximation \eqref{Phi}. The red dots highlights location of the semi-classical fixed point \eqref{semicl} . The white lines shows the brunch-cuts of the effective potential \eqref{Phi}. The system's parameters are set to $G=10, \Delta=7,\eta=1$.}
\label{Fig1}
\end{figure}

Further, we will consider only the minus solution as long as we are interested in the behavior of the system near the minimums and saddle point of the "effective potential". We also drop all slowly varying terms and constants and obtain the following approximation for the Wigner function:   
\begin{equation}\label{WignerPot}
    W_0(\alpha,\alpha^*)\approx\exp\left[-\Phi(\alpha,\alpha^*)\right],
\end{equation}
where we introduce the effective potential $\Phi$:
\begin{equation}\label{Phi}
\begin{split}
 &\Phi(\alpha,\alpha^*)=2\alpha^*\alpha+\\
 &\frac{2}{\eta}\mathrm{Im}\left[ \sqrt{\Delta ^2-4 \eta  G\alpha ^2 }-\Delta  \ln \left( \frac{\sqrt{\Delta ^2-4 \eta G\alpha ^2}+\Delta}{\eta}\right)    \right].   
\end{split}
\end{equation}
It is worth mentioning that \eqref{WignerPot} is a good approximation for exact solution only for the first and third quadrants, as shown in Fig. \ref{Fig1}.

In summary, this Section examines the stationary Wigner function for a detuned quantum cavity subjected to two-photon pump and two-photon dissipation. The system obeys the hidden time-symmetry \cite{Roberts2021}, which constrains the form of the stationary Wigner function. This restriction allows us to move from Eq. \eqref{WignerEq} which contains the third-order derivatives to the second-order equation \eqref{difeq1}. This mathematical simplification permits an exact closed-form solution for the Wigner function, expressed in terms of a Hypergeometric function. Although equivalent solutions have previously been reported using P‑representation techniques \cite{Bartolo2016,Minganti2016}  and the quantum absorber method \cite{Roberts2020}, this is the first time they have been directly calculated using the Wigner equation of motion. Building upon this result, we further employ the WKB approximation to derive an effective potential in a compact analytical form. As will be demonstrated in the following Section, this effective potential has a fundamental connection with the real-time instanton formalism, offering a bridge between steady-state properties and dynamical tunneling processes.

\section{Real-time instantons}
\begin{figure}
\includegraphics[width=0.45\textwidth]{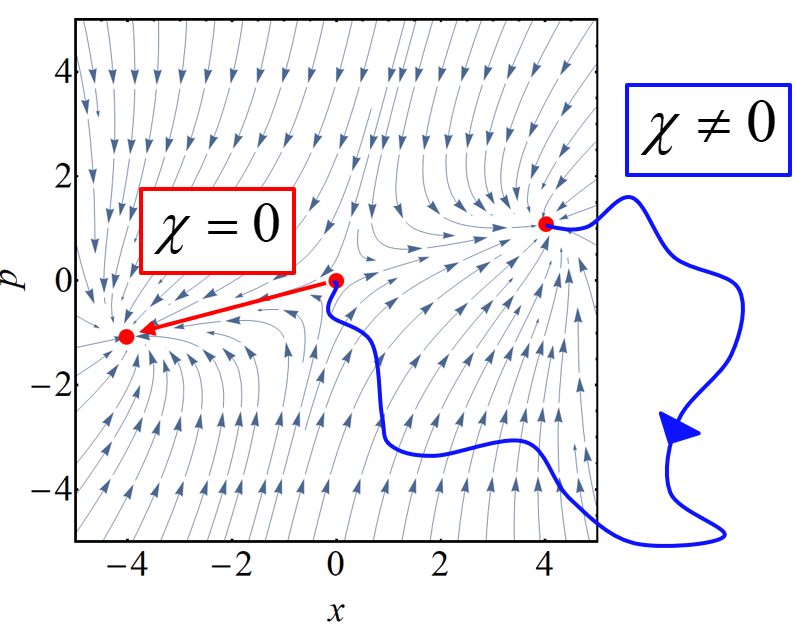}
\caption{The phase portrait of the system in the classical subspace where the quantum field $\chi$ is zero. The instanton trajectory consists of two parts colored blue and red. The blue line schematically illustrates the instanton trajectory, starting from the classical fixed point and ending at the saddle point. Along this trajectory the system explores a quantum four-dimensional phase space where quantum field has nonzero value. The red line shows how the system descends from the saddle point to a second attractive point in the classical subspace.}
\label{Fig2}
\end{figure}

As we examined in a previous Section, the stationary Wigner distribution for the detuned two-photon cavity reveals two distinct, equally populated attractors in phase space (see Fig. \ref{Fig1}). However, if we initialize the system near only one of the fixed points, it will acquire finite life-time due to quantum fluctuations, which can induce switching between the two metastable states with opposite phases \cite{Marthaler2007,Dykman1994,Dykman1998,Dykman2007,Dykman2012}. In this section, we discuss the instanton approach to calculate the decoherence rate, where significant progress has been made in recent years \cite{Thompson2022,Carde2025,Sepulcre2025,Lee2025}. It is worth mentioning the paper by Leon Carde et al. \cite{Carde2025}, where the exponent of the switching rate was calculated for a bistable photonic cavity using the hidden time-reversal symmetry and the P-representation. However, in this Section, we will show how to connect the real-time instanton and Wigner function approaches. Let us start with the Keldysh partition function \cite{Sieberer2016,Kamenev}:
\begin{equation}
  \mathcal{Z} = \int\mathcal{D}[a_{\mathrm{cl}}, a_{\mathrm{q}}]\, e^{i S[a_{\mathrm{cl}}, a_{\mathrm{q}}]},  
\end{equation}
where $\mathcal{D}[a_{\mathrm{cl}}, a_{\mathrm{q}}]$ is a standard functional measure and $S$ is a Keldysh action: 
\begin{equation}\label{action}
    S[\alpha_{cl},\alpha_{q}]=\int dt\left( \bar{\alpha}_{q}i\p_t \alpha_{cl}+\bar{\alpha}_{cl}i\p_t \alpha_{q}-i \mathcal{L}\right),
\end{equation}
where $\alpha_{cl}/\alpha_{q}$ is a classical/quantum field and $\mathcal{L}$ is a Liouvillian density functional, which is given in Appendix \ref{AppEqofM}.
We will consider a saddle-point approximation to the partition functions described above. In this approximation, the main contribution to the time dynamics comes from the one-instanton process, which leads to an exponential dependence of the decoherence rate \cite{Kamenev,Thompson2022}:
\begin{equation}\label{GammaAction}
   \Gamma \propto \exp(i S),
\end{equation}
where $S$ is an action evaluated along the trajectory that extremizes it. To calculate the instanton trajectories, one must find the following classical equations of motion: 
\begin{equation}\label{EqofM}
\begin{split}
       & \p_t \alpha_{cl}=\p \mathcal{L}/\p\bar{\alpha}_q;
       \quad\p_t\alpha_q=\p\mathcal{L}/\p\bar{\alpha}_{cl}\\
       & \p_t \bar{\alpha}_{cl}=-\p\mathcal{L}/\p \alpha_q;
       \quad\p_t\bar{\alpha}_{q}=-\p\mathcal{L}/\p \alpha_{cl},
\end{split}
\end{equation}
where the functional $\mathcal{L}$ is a conserved quantity ($\p_t \mathcal{L}=0$) and plays the role of an effective Hamiltonian that generates the equations of motion. Nominally, $\bar{\alpha}_{cl/q}$ is the complex conjugated to $\alpha_{cl/q}$, respectively. However, this structure is absent in the saddle-point equations \eqref{EqofM} (for details, see the Appendix \ref{AppEqofM}). To restore this feature, one can rename the photonic fields in the following way:
\begin{equation}
\begin{gathered}
\alpha_{cl} = \sqrt{2}\,\alpha,\quad
\bar{\alpha}_{cl} = \sqrt{2}\,\bar{\alpha},\\
\alpha_q = -\sqrt{2}\,\bar{\chi},\quad
\bar{\alpha}_q = \sqrt{2}\,\chi,
\end{gathered}
\end{equation}
where $\bar{\chi}=\chi^*$ and $\bar{\alpha}=\alpha^*$. As a result, we obtain the following equations of motion for the variables $\chi$ and $\alpha$:
\begin{equation}\label{Eqofmalpchi}
\begin{aligned}
& \partial_t \alpha = 
i\Delta\alpha+G \bar{\alpha}
-\eta\left(\bar{\alpha} (\alpha^2+\bar{\chi}^2)
+2 \bar{\chi} \chi \alpha
-4 \bar{\chi}\bar{\alpha}\alpha\right), \\
& \partial_t \chi = 
-i\Delta\chi-G \bar{\chi}
+\eta\left(\bar{\chi} (\chi^2+\bar{\alpha}^2)
+2 \bar{\alpha} \alpha \chi
-4 \bar{\alpha}\bar{\chi}\chi\right),
\end{aligned}
\end{equation}
where the dynamics of photonic variables $\partial_t \alpha = \partial_\chi L,\partial_t \chi = -\partial_\alpha L$ is generated by the effective Hamiltonian $L$:
\begin{equation}
\begin{aligned}
 & L= 
i\Delta(\alpha\chi 
-\bar{\alpha}\bar{\chi})
+ G(\bar{\alpha}\,\chi
+ \alpha\,\bar{\chi})-\\
& - \eta\left(
\bar{\alpha}\alpha\alpha\chi
+ \bar{\chi}\bar{\alpha}\bar{\alpha}\alpha
+ \bar{\chi}\chi\chi\alpha
+ \bar{\alpha}\bar{\chi}\bar{\chi}\chi
- 4\,\bar{\alpha}\alpha\bar{\chi}\chi
\right).
\end{aligned}
\end{equation}
The introduced photonic variables $\alpha, \chi$ have a simple physical meaning: $\alpha$ is a classical field and $\chi$ is a quantum field. It also follows from equations \eqref{Eqofmalpchi} that the quantum field $\chi$ is a time-reversed version of the classical field $\alpha$. The instanton motion takes place in the four real-dimensional phase space, since $\alpha$ and $\chi$ are complex variables. If quantum fluctuations are neglected ($\chi=0$), the first equation in \eqref{Eqofmalpchi} transforms into the semi-classical equation \cite{Bartolo2016}, which takes the following form:
\begin{equation}\label{alpha}
 \p_t \alpha=i  \Delta \alpha  +  G\alpha^*-\eta \alpha ^2 \alpha^* .
\end{equation}
The stationary solutions of this equation \eqref{alpha}, which satisfy the condition $\partial_t \alpha = 0$, have several fixed points in the phase diagram: two stable points $\alpha = \pm \alpha_0$ and one unstable point $\alpha_{\rm{saddle}}$, which has a saddle-point character (Fig. \ref{Fig2}). All three points are well separated in phase space.
 \begin{equation}\label{semicl}
 \begin{gathered}
 \alpha_0 = \sqrt{\frac{\sqrt{G^2 - \Delta^2}}{\eta}} \cdot \sqrt{\frac{\sqrt{G^2 - \Delta^2} + i \Delta}{G}}, \\
 \alpha_{\text{saddle}} = 0.
 \end{gathered}
 \end{equation}
Thus, there is a classical two-dimensional subspace ($\chi=0$) of the full four-dimensional phase space, where the classical points live, as shown in Fig. \ref{Fig2}. However, they become metastable when the whole quantum phase space is taken into account. In Fig. \ref{Fig2} one can see that the quantum fluctuations can push the system out of the classical subspace, where the quantum field becomes nonzero ($\chi\neq0$ and the blue line). After traveling across the quantum phase space, the system can return to the classical subspace. Then it descends from a saddle point to a second point of attraction, as depicted by the red line in Fig.  \ref{Fig2}. 
The discussed instanton trajectory thus starts within the classical subspace, where the quantum variable and the effective Hamiltonian $L$ are zero. However, the effective Hamiltonian, being the conserved quantity, remains equal to zero along the entire instanton trajectory:
\begin{equation}\label{Lzero}
    L[\alpha,\bar{\alpha},\chi,\bar{\chi}] = 0.
\end{equation}
When frequency detuning is zero ($\Delta=0$) one can consider the photonic fields as real variables \cite{Thompson2022}. In this situation, the full quantum phase space shrinks to two dimensions with a one-dimensional classical subspace. This allows one to map the problem of real-time instanton motion to the classical stochastic dynamics with one degree of freedom. In this limiting case, one can easily find the dependence of the quantum field on the classical one explicitly from Eq. \eqref{Lzero} and obtain the following: 
\begin{equation}\label{Chi1}
   \chi = \alpha\pm\sqrt{G/\eta}.
\end{equation}
Using Eq. \eqref{Chi1} one can calculate the action along the instanton trajectory and find the decoherence rate \eqref{GammaAction} \cite{Thompson2022}:
\begin{equation}
   \Gamma \propto\exp(-2G/\eta).
\end{equation}
\begin{figure}
\includegraphics[width=0.43\textwidth]{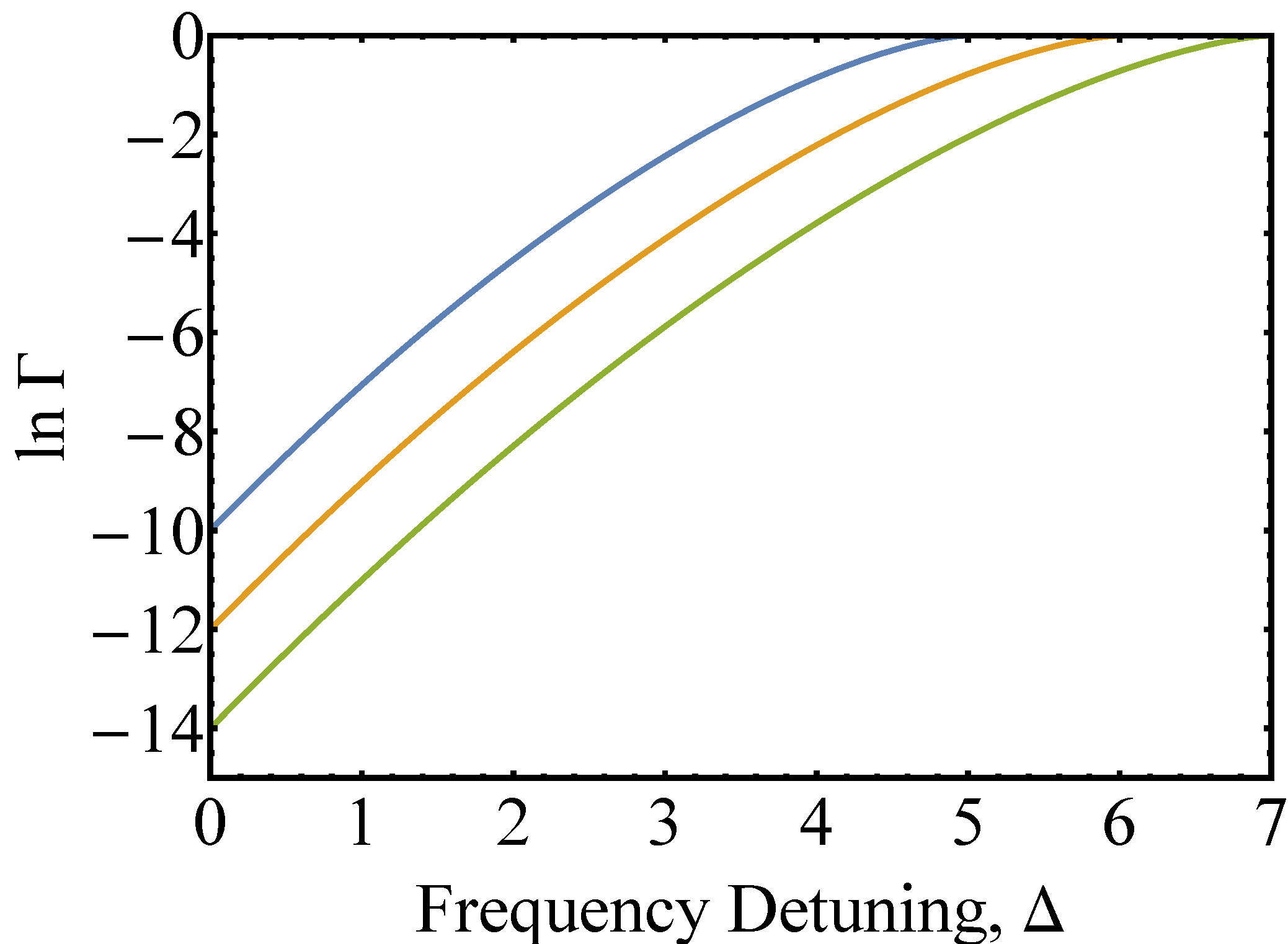}
\caption{The dependence of the logarithm of the switching rate on the frequency detuning $\Delta$, plotted for different values of two-photon pump: $G=7$ (green curve), $G=6$ (orange curve), $G=5$ (blue curve), and for the two-photon pumping rate parameter $\eta=1$.}
\label{Fig3}
\end{figure}
However, the nonzero value of the detuning returns the photonic variables to a complex space, greatly complicating the solution of Eq.  \eqref{Lzero}.  However, correctly guessing the form of the solution can help overcome this difficulty. Let us search for it as given: 
\begin{equation}\label{chivarphi}
\begin{split}
\chi = \bar{\alpha} + f(\alpha),\quad\bar{\chi} = \alpha + \bar{f}(\bar{\alpha}).
\end{split}
\end{equation}
The main simplification comes from the fact that the variable $f$ depends only on $\alpha$, rather than being the function of the two variables $\alpha$ and $\bar{\alpha}$. This trick helped us in the previous Section, where the decomposition of the stationary Wigner function is proposed in the same way \eqref{Decomp}. Substituting the expression \eqref{chivarphi} into \eqref{Lzero}, we obtain the following equation:
\begin{equation}\label{A1A2A3}
A_1(\alpha) + A_2(\bar{\alpha}) + A_3(\alpha,\bar{\alpha}) = 0,
\end{equation}
where three different terms arise:
\begin{equation}
\begin{split}
&A_1(\alpha) = 
\alpha\left(i\Delta f 
+ \alpha\left(G - \eta f^2\right)\right),\\
 &A_2(\bar{\alpha})= 
\bar{\alpha}\left(-i\Delta \bar{f} 
+ \bar{\alpha}\left(G - \eta \bar{f}^2\right)\right),\\
 &A_3(\alpha,\bar{\alpha})= 
\bar{f}\,\alpha\left(G - \eta f^2\right)
+ \bar{\alpha}\,f\left(G - \eta \bar{f}^2\right).
\end{split}
\end{equation}
Thus, the expression \eqref{A1A2A3} has the same structure that we discussed in the previous Section (see Appendix \ref{AppPsi} for details). Following this established logic, we first solve the following equations:
\[
A_1(\alpha)=0,\qquad A_2(\bar{\alpha})=0,
\]
on functions $f$ and $\bar{f}$ and obtain:
\begin{equation}\label{f}
\begin{split}
&f_{\pm}(\alpha) = 
i\frac{\Delta \pm \sqrt{\Delta^2 - 4 G \eta\, \alpha^2}}
{2\eta\,\alpha},\\
&\bar{f}_{\pm}(\bar{\alpha}) = 
-i\frac{\Delta \pm \sqrt{\Delta^2 - 4 G \eta\, \bar{\alpha}^2}}
{2\eta\,\bar{\alpha}}.
\end{split}
\end{equation}
Substituting \eqref{f} into $A_3(\alpha,\bar{\alpha})$ causes the latter to vanish, thus showing that the functions \eqref{f} satisfy the condition $L = 0$. Further, we will consider only the "minus" solution in Eq. \eqref{f} because it is regular near the saddle point $\alpha=0$. As a result, we have the following expression for the quantum field:
\begin{equation}\label{chifinal}
\begin{split}
\chi_{} &= 
\bar{\alpha} +i\,\frac{\Delta - \sqrt{\Delta^2 - 4 G \eta\, \alpha^2}}
{2\eta\,\alpha},
\end{split}
\end{equation}
and $\bar{\chi}_{}=\chi_{}^*$. It is easy to check that the quantum field is zero when we set $\alpha=\pm\alpha_0$ or $\alpha=0$, and the trajectory \eqref{chifinal} starts and ends in the classical subspace. It is also convenient to connect \eqref{chifinal} with effective potential \eqref{Phi} introduced in a previous Section:
\begin{equation}\label{chiandPhi}
\begin{split}
\chi_{} =\frac{1}{2}\frac{\p\Phi(\alpha,\bar{\alpha})}{\p \alpha},\quad\bar{\chi}_{} =\frac{1}{2}\frac{\p\Phi(\alpha,\bar{\alpha})}{\p \bar{\alpha}}.
\end{split}
\end{equation}
Eq. \eqref{chiandPhi} is a major result of this paper highlighting the deep connection between the Wigner representation and the real-time instantons. The next step is a calculation of the decoherence rate \ref{GammaAction}. This requires calculating the instanton action:
\begin{equation}\label{InstAct}
    i S=-2\int_{t_i}^{tf} dt (\chi\p_t \alpha-\bar{\alpha}\p_t\bar{\chi}),
\end{equation}
where part of the action is discarded because along the instanton trajectory $\mathcal{L}=0$. 

As we mentioned earlier, the instanton trajectory consists of two contributions. The first part of the trajectory begins at the attractive fixed point and ends at the saddle point (blue line in Fig. \ref{Fig2}). The second part starts at the saddle point and proceeds to the other attractive fixed point (red line in Fig. \ref{Fig2}). However, the second part of the trajectory does not give the contribution to the instanton action \eqref{InstAct} because the quantum field and its time derivative are equal zero. 
After integration by part and utilization of Eq. \eqref{chiandPhi} we obtain the following simple relation for the decoherence rate and  instanton action: 
\begin{equation}\label{lnGamma}
\begin{gathered}
   \ln{\Gamma} \approx iS= -\Phi(\alpha,\alpha^*)|_{\alpha=\alpha_{\text{saddle}}}^{\alpha=\alpha_0}=\\=-2\frac{\sqrt{G^2-\Delta^2}}{\eta}+2\frac{\Delta}{\eta}\arctan\left(\frac{\sqrt{G^2-\Delta^2}}{\Delta}\right),
\end{gathered}
\end{equation}
where we substitute the attractive/saddle point $\alpha_{0/\text{saddle}}$ \eqref{semicl} into the effective potential $\Phi$ \eqref{Phi}. The independence of the action from from the exact shape of the instanton trajectory is a manifestation of quantum detailed balance, i.e., a hidden time-reversal symmetry \cite{Roberts2021}. 

In Fig.~\ref{Fig3} we show the dependence of \eqref{lnGamma} on frequency detuning $\Delta$. When detuning tends to zero, \eqref{lnGamma} tends to $-2G/\eta$, which is consistent with the results reported in~\cite{Thompson2022}. One can see that  \eqref{lnGamma} grows with increasing detuning, which suppress quantum activation between meta-stable states. This behavior continues until the critical point $\Delta = G$, where the dissipative phase transition occurs~\cite{Mylnikov2025, Mylnikov2022, Bartolo2016}. Near this point, the logarithm of the decoherence rate takes the form:
\begin{equation}
    \ln \Gamma \approx -\frac{4\sqrt{2}\,(G-\Delta)^{3/2}}{3\eta\sqrt{G}},
\end{equation}
and tends to zero as a power law. This is related to the fact that $\ln \Gamma$ acts as an effective potential barrier. At the critical point $\Delta=G$, two fixed points merge, the barrier disappears, and the quantum activation process ceases to exist.
\section{Conclusion}
In this work, we have investigated the quantum dynamics and steady-state properties of a driven-dissipative microwave cavity subject to the two-photon pump and two-photon dissipation. The system exhibits quantum bistability, characterized by multiple steady states of the photonic field. Although stable in the semiclassical limit, these states acquire a nontrivial Wigner distribution function and finite lifetime when quantum fluctuations are taken into account. The key outcome of our analysis is the establishment of a deep and explicit connection between the Wigner function and the real-time instanton formalism describing quantum activation process between metastable states. By leveraging a hidden time-reversal symmetry inherent to the detuned two-photon cavity, we derived an exact factorized form of the stationary Wigner distribution. In the regime of weak nonlinear dissipation, this result simplifies further into an effective potential, which compactly encodes the nontrivial phase-space structure of the steady-state. These results have immediate application in the framework of the real-time instantons describing quantum dynamics of the system. It is governed by a pair of photonic fields, classical and quantum, which are connected via time-reversal symmetry.  We demonstrated that the quantum field along an instanton (a saddle trajectory in the extended four-dimensional quantum phase space) is directly related to the gradient of the introduced effective potential. This relation bridges the steady-state quantum properties with real-time quantum activation dynamics. From the instanton analysis, we also obtained an analytical expression for the quantum decoherence rate between the metastable states. This rate is directly determined by the Wigner effective potential, providing a unified description of metastability and decoherence. In summary, this work unifies several theoretical techniques: the Wigner phase-space methods, hidden time reversal symmetry, the Keldysh field theory, and instanton calculus into a coherent framework for studying quantum metastability in driven-dissipative nonlinear resonators. Our results highlight how quantum fluctuations, captured via the Wigner function, govern the lifetime of macroscopic states of the system and enable rare tunneling events that can be described as real-time instantons. These results are valuable for optimizing the design and performance of bosonic cat code architectures.

\begin{acknowledgments}
We are grateful to Mark Dykman and Foster Thompson for valuable discussions. S.O. Potashin thanks the Russian Science Foundation (Project No. 25-12-00212) for financial support of the theoretical study of the Wigner function. V. Yu. Mylnikov acknowledges the support of numerical simulations and the instanton approach by the Foundation for the Advancement of Theoretical Physics and Mathematics “BASIS.” 
\end{acknowledgments}

\appendix

\section{Factorization of the Wigner function 
}\label{AppPsi}

In Section III, the stationary solution of the Wigner equation \eqref{WignerEq} is sought in the form of Eq.  \eqref{Decomp}. After its substitution to Eq. \eqref{WignerEq}, we can obtain the following equation for functions $\Psi_1$ and $\Psi_2$: 
\begin{widetext}
\begin{equation}\label{Decomp2}
\begin{gathered}
-2 \alpha  \Psi_2(\alpha^* ) \left(\alpha  \eta  \Psi_1''(\alpha )+2 i \Delta  \Psi_1'(\alpha )-4 \alpha  G \Psi_1(\alpha )\right) -\\-2 \alpha^*   \Psi_1(\alpha ) \left(\alpha^*   \eta  \Psi_2''(\alpha^* )-2 i \Delta  \Psi_2'(\alpha^*  )-4 \alpha^*   G \Psi_2(\alpha^*  )\right) +\\+\alpha  \Psi_2'(\alpha^* ) \left(\eta  \Psi_1''(\alpha )-4 G \Psi_1(\alpha )\right)+\alpha^*  \Psi_1'(\alpha ) \left(\eta  \Psi_2''(\alpha^* )-4 G \Psi_2(\alpha^* )\right) = 0.
\end{gathered}
\end{equation}
\end{widetext}
where we denote $d\Psi_1(\alpha)/d\alpha=\Psi'_1(\alpha)$. In this equation, we can identify contributions with different meanings. We will refer to the first contribution as the expression in the first line-the bracket preceding $-2\alpha\Psi_2(\alpha^*)$, which depends only on $\Psi_1(\alpha)$. The second contribution will be taken as the expression in the second line-the bracket preceding $2\alpha^*\Psi_1(\alpha)$, which depends only on $\Psi_2(\alpha^*)$. The third contribution (the third line of equation \eqref{Decomp2}) depends simultaneously on both $\Psi_1(\alpha)$ and $\Psi_2(\alpha^*)$.

It is natural to require that the first and second contributions vanish independently. If this holds, then the third contribution will automatically become zero. Thus, from equation \eqref{Decomp2} we proceed to the simpler equations \eqref{difeq1} and \eqref{difeq2}. 
\section{WKB method}\label{AppWKB}
This section provides a detailed calculation of the function $\Psi_1(\alpha)$ using the WKB method. We begin the analysis with equation \eqref{difeq1}:
\begin{equation}\label{difeq11}
    \alpha \eta \Psi_1''(\alpha)
    + 2 i \Delta \Psi_1'(\alpha)
    - 4 \alpha G \Psi_1(\alpha) = 0.
\end{equation}
A formal solution to this second-order differential equation for the function $\Psi_1(\alpha)$ can be conveniently sought in exponential form:
\begin{equation}
    \Psi_1(z) = \exp\left(\frac{\phi(z)}{\eta}\right),
\end{equation}
where we introduce $z=\alpha \sqrt{\eta}$.  Substituting this expression into the original equation \eqref{difeq11} we obtain the following equation:
\begin{equation}\label{difeq}
    -4 G z + 2 i \Delta \phi' + z (\phi')^2 + z \eta \phi'' = 0,
\end{equation}
where $\phi' = d\phi/dz$. Let us solve this equation \eqref{difeq} using the WKB method. We expand the function $\phi$ in powers of the two-photon dissipation rate $\eta$:
\begin{equation}
    \phi(z) \approx \phi_0(z) + \eta \phi_1(z).
\end{equation}
One can substitute this expansion into the equation \eqref{difeq} and obtain:
\begin{equation}\label{phi0}
    \bigl[-4 G z + 2 i \Delta \phi_0' + z (\phi_0')^2\bigr]
    + \eta\bigl(2 i \Delta \phi_1' + 2 z \phi_0' \phi_1' + z \phi_0''\bigr) = 0.
\end{equation}
First, we set the expression in square brackets in equation \eqref{phi0} to zero and obtain a quadratic equation for $\phi_0'$, the solution of which is:
\begin{equation}
    \phi_{0,\pm}'=\frac{-i \Delta\mp\sqrt{\Delta^2-4 G z^2}}{z}.
\end{equation}
Integrating over the variable $z$, we find the explicit form of the functions $\phi_{0,\pm}$, which leads to the expression \eqref{phi0pm}.

Let us now proceed the calculation of the function $\phi_{1,\pm}$. Consider the expression in parentheses in equation \eqref{phi0}. Setting it to zero, it can be shown that $\phi_{1,\pm}'$ is expressed in terms of $\phi_{0,\pm}'$ as follows:
\begin{equation}\label{phi1}
    \phi_{1,\pm}' = -\frac{z\, \phi_{0,\pm}''}{2\bigl(i\Delta + z \phi_{0,\pm}'\bigr)}.
\end{equation}
First, we choose the function $\phi_{0,+}$ and compute the corresponding expression for $\phi_{1,+}$. Substituting the previously found expression for $\phi_{0,+}'$ into \eqref{phi1}, we get:
\begin{equation}
\begin{split}
    &\phi_{1,+}(z)
    = \int d z\,
    \frac{\Delta\bigl(-\Delta + \sqrt{\Delta^2 - 4 G z^2}\bigr)}
    {8 G z^3 - 2 \Delta^2 z} \\
    &= \frac{1}{2}\ln\left[
    \frac{\Delta + \sqrt{\Delta^2 - 4 G z^2}}
    { \sqrt{\Delta^2 - 4 G z^2}}
    \right].
\end{split}
\end{equation}
Now consider the second case, corresponding to $\phi_{0,-}$. Then the expression for $\phi_{1,-}$ becomes:
\begin{equation}
\begin{split}
    &\phi_{1,-}(z)
    = \int d z\,
    \frac{\Delta\bigl(-\Delta - \sqrt{\Delta^2 - 4 G z^2}\bigr)}
    {8 G z^3 - 2 \Delta^2 z} \\
    &= -\frac{1}{2}\ln\left[
    \frac{\sqrt{\Delta^2 - 4 G z^2}\bigl(\Delta + \sqrt{\Delta^2 - 4 G z}\bigr)}
    {-4 G z^2}
    \right].
\end{split}
\end{equation}
For convenience, we introduce the amplitude $A_{\pm}$, which is related to the function $\phi_{1,\pm}(z)$ as follows:
\begin{equation}\label{Aplus}
A_{+}=\exp{\left(\phi_{1,+}(z)\right)}=\sqrt{\frac{\Delta+\sqrt{\Delta^2-4 G z^2}}{\sqrt{\Delta^2-4 G z ^2}}} 
  \end{equation}
\begin{equation}\label{Aminus}
\begin{split}
   & A_{-}=\exp{\left(\phi_{1,+}(z)\right)}=\\&=i\sqrt{
    \frac{4 G z^2}{\sqrt{\Delta^2 - 4 G z^2}\bigl(\Delta + \sqrt{\Delta^2 - 4 G z}\bigr)}}
\end{split} 
\end{equation}

Thus, having found the functions $\phi_{0,\pm}$ and $\phi_{1,\pm}$, we obtain the explicit form of the function $\Psi_1(\alpha)$. 
\section{Asymptotics of the $\Psi_1$ function  
}\label{App2Sol}
Let us introduce $\delta=\Delta/\eta$ and  $g=G/\eta$ and rewrite \eqref{PsiExact} in the following form:
\begin{equation}\label{PsiApp}
    \Psi_1(\alpha) =\mathcal{N} \exp\left({2 
   \sqrt{g}\alpha}\right) \, _1F_1\left(i\delta;2i\delta;-4 \sqrt{g} \alpha 
  \right).
\end{equation}
One can use the asymptotics of Kummer's hypergeometric function for large arguments \cite{Gradshteyn}: 
\begin{equation}
_1F_1\left(a;b;z\right)\approx
\frac{\Gamma (b)}{\Gamma (b-a)}e^{-a\log(-z)}+
\frac{\Gamma (b)}{\Gamma (a)}e^{z+(a-b)\log(z)}.
\end{equation}
and find the asymptotics of the function \eqref{PsiApp}: 
\begin{equation}
\begin{gathered}
\Psi_1(\alpha) \approx\\
\approx\mathcal{N} \frac{\Gamma (2i\delta)}{\Gamma (i\delta)} e^{-i\delta\log(4 \sqrt{g} \alpha)}
\left(e^{2\sqrt{g} \alpha }+e^{-2 \sqrt{g} \alpha-\pi\delta }\right).
\end{gathered}
\end{equation}
We can also find the asymptotics of the WKB solutions for large $\alpha$: 
\begin{equation}
    \begin{split}
        &   \Psi_{1}^\pm\approx\exp{\left(\mp2 \sqrt{g}\alpha-i\delta \ln{(2\sqrt{g}\alpha)-\pi \delta/2}\right)}.
    \end{split}
\end{equation}
As a result, we find the expansion coefficients for $\Psi_1(\alpha)$ by representing this function as a superposition of WKB solutions:
\begin{equation}
    \Psi_1(\alpha)=C_{+}\Psi^{+}_{1}+C_{-}\Psi^{-}_{1},
\end{equation}
where $$C_{\pm}= \mathcal{N} 2^{-i\delta} \frac{\Gamma (2i\delta)}{\Gamma (i\delta)}\exp{\left(\mp\frac{\delta\pi}{2}\right)}.$$
\section{Equations of motion}
\label{AppEqofM}
In this Appendix, we write the exact form of the Liouvillian density functional:

\begin{widetext}
\begin{equation}\label{Li}
\begin{split}
      \mathcal{L}=i\Delta (\alpha_{cl}\bar{\alpha}_{q}+\alpha_{q}\bar{\alpha}_{cl})+G (\bar{\alpha}_{cl}\bar{\alpha}_{q}-\alpha_{q}\alpha_{cl})-
      \frac{\eta}{2}\left(\alpha_{cl}^2\bar{\alpha}_{cl}\bar{\alpha}_{q}-\bar{\alpha}_{cl}^2\alpha_{cl}\alpha_{q}+\alpha_{q}^2\bar{\alpha}_{cl}\bar{\alpha}_{q}-\bar{\alpha}_{q}^2\alpha_{cl}\alpha_{q}+4\bar{\alpha}_{cl}\alpha_{cl}\bar{\alpha}_{q}\alpha_{q}\right).
\end{split}
\end{equation}
\end{widetext}
After variation of the Keldysh action \eqref{action}, one can obtain the explicit form of the equations of motion \eqref{EqofM}:
\begin{widetext}
\begin{equation}\label{EqofM2}
\begin{split}
& \partial_t \alpha_{cl}=i\Delta\alpha_{cl}+G \bar{\alpha}_{cl}-\frac{\eta}{2}\left(\bar{\alpha}_{cl} \alpha_{cl}^2+\bar{\alpha}_{cl} \alpha_q^2-2 \bar{\alpha}_q \alpha_q \alpha_{cl}+4 \bar{\alpha}_{cl} \alpha_{cl} \alpha_q\right), \\
& \partial_t \bar{\alpha}_{cl}=-i\Delta\bar{\alpha}_{cl}+G \alpha_{cl}-\frac{\eta}{2}\left(\alpha_{cl} \bar{\alpha}_{cl}^2+\alpha_{cl} \bar{\alpha}_q^2-2 \alpha_q \bar{\alpha}_q \bar{\alpha}_{cl}-4 \alpha_{cl} \bar{\alpha}_{cl} \bar{\alpha}_q\right), \\
& \partial_t \alpha_q=-i\Delta\alpha_{q}+G \bar{\alpha}_q-\frac{\eta}{2}\left(\bar{\alpha}_q \alpha_{cl}^2+\bar{\alpha}_q \alpha_q^2-2 \bar{\alpha}_{cl} \alpha_q \alpha_{cl}+4 \bar{\alpha}_q \alpha_{cl} \alpha_q\right), \\
& \partial_t \bar{\alpha}_q=i\Delta\bar{\alpha}_q+G \alpha_q-\frac{\eta}{2}\left(\alpha_q \bar{\alpha}_{cl}^2+\alpha_q \bar{\alpha}_q^2-2 \alpha_{cl} \bar{\alpha}_q \bar{\alpha}_{cl}-4 \alpha_q \bar{\alpha}_{cl} \bar{\alpha}_q\right).
\end{split}
\end{equation}
\end{widetext}
These equations are necessary for calculating the instanton trajectory. Note that the conjugacy between the components $\alpha_{cl}$ and $\bar{\alpha}_{cl}$, as well as $\alpha_q$ and $\bar{\alpha}_q$, is violated due to the latter contribution to the equations of motion \eqref{EqofM2}.

\bibliography{main}

\end{document}